\documentstyle[preprint,aps]{revtex}
\tighten
\draft
\begin{document}
\widetext
\preprint{CLNS 96/1412}
\bigskip
\bigskip
\title{Three-Family $SO(10)$ Grand Unification in String Theory}
\medskip
\author{Zurab Kakushadze\cite{foot1} and S.-H. Henry Tye}
\bigskip
\address{Newman Laboratory of Nuclear Studies,
Cornell University,
Ithaca, NY 14853-5001, USA}
\date{May 30, 1996}
\bigskip
\medskip
\maketitle
\begin{abstract}	
{}The construction of a supersymmetric $SO(10)$ grand unification 
with $5$ left-handed and $2$ right-handed families
in the four-dimensional heterotic string theory is presented. The
model has one $SO(10)$ adjoint Higgs field. The $SO(10)$ current algebra 
is realized at level $3$.
\end{abstract}
\narrowtext

{}The outstanding question of superstring theory is how does it
describe our universe. The space of classical supersymmetric string 
vacua has a large degeneracy, which may be described by a 
set of parameters, or moduli. We expect string 
dynamics to lift, partially or completely, 
this huge degeneracy in the moduli space.
Each point in this moduli space corresponds to a particular string model. 
To be specific, let us consider the heterotic superstring case.
Ignoring the possibility of enhanced gauge symmetry from
non-perturbative effects, the rank of the gauge group is 22 or less 
(for our counting purpose, $U(1)$ has rank 1). After accommodating
the standard model of strong and electroweak interactions (with 
minimum rank 4), there is still plenty of room 
({\em i.e.}, with maximum rank 18) for a large hidden sector. The
possible choices of the hidden sector
are myriad and largely unexplored, and the dynamics in each case
is very complicated. The difficulty of string phenomenology 
is the lack of an objective criteria that would select a particular 
model among the numerous possibilities; 
that is, the moduli space of the hidden sector is
too big for the string dynamics to be analyzed systematically.
This difficulty may be solved by considering grand unification in 
string theory. As we shall see, after imposing some rather simple 
phenomenological constraints, the hidden sector seems to be 
unique and the remaining moduli space is essentially reduced 
to a one-dimensional space. The string dynamics in this case
should be within reach. 

{}The apparent unification of the gauge couplings in the context of
supersymmetry when extrapolated to high energy scales has created a lot
of interest in supersymmetric grand unified theories. To realize a
grand unified model in the superstring theory, there is a very
stringent contraint. It is well known that, in field theory, adjoint Higgs
(or other appropriate higher dimensional) representation is 
necessary for a grand unified
gauge group to break spontaneously to the 
$SU(3) \otimes SU(2) \otimes U(1)$ gauge group of the standard model.
It is also known that, for current algebras at level 1, space-time
supersymmetry with chiral fermions do not co-exist with massless
scalar fields in the adjoint or higher dimensional representations of
the gauge group in heterotic string models. From 
these facts, one concludes that a grand unified model in the
superstring theory is possible only if the grand unified gauge
group comes from current algebras at levels higher than 1.

{}Grand unified models in the
superstring theory are sometimes referred to as
grand unified string theories (GUST). The first GUST analysis was given by
Lewellen\cite{lew}. In particular, he constructed an $SO(10)$ GUST with four
chiral families. Next, Schwartz extended the construction to include
an $SU(5)$ GUST, also with four chiral families\cite{schw}. More
recently, Erler used the orbifold method\cite{orb} to construct 
$E_6$ GUSTs, again with an even number of families\cite{erler}. In the 
mean time, there are a number of other interesting related 
works\cite{try}.
Since nature seems to have only three light families, attempts
were made to construct a GUST with three chiral families, so far,
unsuccessful. This suggests that a 3-family-GUST either does not exist, or,
more interestingly, is extremely limited.
The GUST models with even number of families mentioned above
all involve a level-2 gauge group. So, to find a GUST with three families,
it is natural to go to level-3 gauge groups. 
In this work, we shall report the construction of such a model: 
An $N=1$ supersymmetric $SO(10)$ GUST with an adjoint Higgs and 
three chiral families; to be more precise, the model has five 
left-handed and two right-handed chiral families.
        
{}The model has the gauge group
$SU(2)_1 \otimes U(1) \otimes M \otimes SO(10)_3 \otimes U(1)$,
where the subscripts indicate the level of the current algebra.
Here, $M$ can be $SU(2)_3 \otimes U(1)$, $U(1)^2$, $U(1)$, or empty. 
{}There is only one adjoint Higgs representation in the $SO(10)$.
Each of the five ${\bf 16}_L$ and the two ${\bf 16}_R$ families of 
$SO(10)$ is accompanied by a ${\bf 10}$ and a singlet of $SO(10)$. 
They have quantum numbers
in $U(1)^2 \otimes M$, but are singlets under the $SU(2)_1$.
So, by definition, the $SU(2)_1$ gauge group is the hidden sector,
while $M$ may be considered as a horizontal symmetry.
There are massless supermultiplets that form doublets in this hidden 
$SU(2)_1$. They are singlets under $SO(10)$ and neutral under the $U(1)s$.
However, these doublets, as well as the chiral families of $SO(10)$, 
have non-trivial quantum values in $M$. 
So we may also consider $M$ as the gauge group for the
messenger/mediator sector, linking the hidden and the visible sectors.

{}Phenomenologically, one wants a hidden sector that will become 
strong so that supersymmetry may be dynamically broken.
The gauge coupling of a given group $G$ in the model at a
scale $\mu$ below the string scale $m_s$ is related to it via:
\begin{equation}
 1/\alpha_G(\mu)= k_G/\alpha_{\mbox{string}} 
 + ({b_0/{4\pi}}) \ln ({m_s^2/ \mu^2})~,
\end{equation}
where $k_G$ is the level of the gauge group. For a $U(1)$ gauge theory,
$1/k=2r^2$ if the $U(1)$ charge is normalized so that the lowest
allowed value is $\pm 1$ (with conformal highest weight $r^2/2$), 
and $r$ is the compactification 
radius of the corresponding world-sheet boson.
The constant $b_0$ is the one-loop coefficient of the beta-function.
The hidden sector $SU(2)_1$ is asymptotically free while $M$ is not. At the 
string scale, the hidden $SU(2)_1$ coupling $\alpha_2$ is three times that 
of the $SO(10)_3$. So, for typical values of the $SO(10)$ grand unification
coupling, $\alpha_2$ becomes large at a rather low scale, within a few orders
of magnitude above the electroweak scale. If $M$ is empty, the hidden 
sector physics will have negligible impact on the physics in the 
visible sector. Since this is phenomenologically undesirable, we shall 
demand a non-empty $M$ in the $3$-family GUST.

{}In Table I, we give the massless spectra for two of these cases.
The first model is
$SU(2)_1 \otimes SU(2)_3 \otimes SO(10)_3 \otimes U(1)^3$, with
the $U(1)$ charges normalized to radii $(1/\sqrt{6}, 1/3\sqrt{2}, 1/6)$.
The second model is $SU(2)_1 \otimes SO(10)_3 \otimes U(1)^4$, with the
$U(1)$ charges normalized to radii $(1/\sqrt{6},1/6\sqrt{2},1/6\sqrt{2},1/6)$.
It is easy to check that both models are anomaly-free.

{}The spectrum of the $M=U(1)$ model can be obtained in the
same way as for the other two models, which we will discuss in a moment. 
Since these models are connected by flat moduli, one may take an effective
field theory approach to obtain the massless spectrum of the $M=U(1)$ model.
Give one of the $SU(2)_3$ doublets in the $U$ sector of the 
$M=SU(2)_3 \otimes U(1)$ (radius $1/3 \sqrt {2}$) model a 
non-zero expectation value, and the gauge group $SU(2)_3 \otimes U(1)$ 
breaks to $U(1)$, with the assignment of this $U(1)$ charge 
completely determined. In the field theoretic approach, 
we may also use the $SO(10)$ adjoint Higgs fields to 
break $SO(10)$ to $SU(5)$ (with its own adjoint Higgs fields).

{}Let us turn to the construction of the models, which is carried out
in the orbifold framework\cite{orb}.
The construction is achieved by turning on Wilson lines in the
$SO(32)$ model toroidally compactified to four dimensions, followed 
by a ${\bf Z}_6$ orbifold. To make the discussion easier to follow,
we split the ${\bf Z}_6$ twist into 
a ${\bf Z}_3$ twist followed by a ${\bf Z}_2$ twist. 
A more detailed discussion will appear separately \cite{zurab}, where 
other GUSTs are also discussed.

{}Our starting point is an $N=4$
space-time supersymmetric Narain model\cite{narain} with the
momenta of the internal bosons spanning an even self-dual Lorentzian lattice
$\Gamma^{6,22}=\Gamma^{2,2} \otimes \Gamma^{4,4} \otimes \Gamma^{16}$,
where each factor is even self-dual. Here
$\Gamma^{2,2} =\{(p_R \vert\vert p_L ) \}$, 
with $p_R ,p_L \in {\tilde \Gamma}^2$ ($SU(3)$ weight lattice), 
and $p_L - p_R \in \Gamma^2$ ($SU(3)$ root lattice). Note that $\Gamma^2=
\{e_i n^i \}$, and ${\tilde \Gamma}^2 =\{{\tilde e}^i m_i \}$, where $e_i$ are
the $SU(3)$ simple roots, and their duals ${\tilde e}^i$ are the 
corresponding weight
vectors ({\em i.e.}, $e_i \cdot {\tilde e}^j = {\delta_i}^j$, $i,j=1,2$). 
$\Gamma^{16}$ is the self-dual ${\mbox{Spin}}(32)/{\bf Z}_2$ lattice. 
$\Gamma^{4,4}$ is an even 
self-dual Lorentzian lattice that admits a symmetric ${\bf Z}_3$ orbifold 
such that both complex coordinates are simultaneously twisted.
The most general $\Gamma^{4,4}$ that possesses such a
${\bf Z}_3$ symmetry has an $8$-dimensional moduli space and 
a generic gauge group $R=U(1)^4$. After orbifolding, the resulting 
$M$ is empty in this generic case. To obtain a non-empty $M$, 
we restrict ourselves to a special one-dimensional subspace of the 
moduli space, which has an enhanced $R$. 
Recall that $\Gamma^{4,4}$ is a momentum lattice corresponding to
a compactification on a torus defined by $X_I=X_I+E_I$. In our case,
the vectors $E_I$ (and their duals ${\tilde E}^I$) can
be expressed in terms of the $SU(3)$ root and weight vectors
$e_i$ and ${\tilde e}^i$ :
\begin{eqnarray}
 &&E_1 =(e_1 ,0),~E_2 =(e_2 ,0), \\
 &&E_3 =(-h {\tilde e}^2 ,g e_1 ),
 ~E_4 =(h {\tilde e}^1 ,ge_2 ).
\end{eqnarray}
where $g\equiv \sqrt{1-{h^2 /3}}$. 
For $0<h<1$, and with appropriate constant antisymmetric 
background fields, we have an enhanced gauge group
$R=SU(3)\otimes U(1)^2$.
At special points $h=0,1$,
$\Gamma^{4,4}$ can be generated by $( 0 \vert\vert E_I)$ and
$({\tilde E}^I \vert\vert {\tilde E}^I)$, and the gauge symmetry is 
enhanced to $R=SU(3)^2$ and $R=SO(8)$, respectively.
As we shall see, these $3$ cases correspond to
$M=U(1)$, $M=U(1)^2$ and $M=SU(2)_3 \otimes U(1)$, respectively.

{}Next we turn on Wilson lines that break the $SO(32)$
subgroup to $SO(10)^3 \otimes SO(2)$. This must be done in a way so that
the resulting Narain model, which we will refer to as $N1$, still possesses 
the ${\bf Z}_3$ symmetry of the space part of the $N0$ model (so that the
${\bf Z}_3$ orbifolding performed in the next step is possible), and also, 
the three $SO(10)$s must be symmetric under a ${\bf Z}_3$ permutation 
(so that modding out
by this outer automorphism will yield $SO(10)_3$).
The above requirements (up to 
equivalent representations) fix the Wilson lines to have the following form:
\begin{eqnarray}
 &&U_1 =(e_1/2 \vert\vert 0)(P^{(1)}_R/2 \vert\vert
 P^{(1)}_L/2)({\bf s}\vert {\bf 0}\vert {\bf 0}
 \vert {\overline S})~,\\
 &&U_2 =(e_2/2 \vert\vert 0)(P^{(2)}_R/2 \vert\vert
 P^{(2)}_L/2)({\bf 0}\vert {\bf s}\vert {\bf 0}
 \vert {\overline S})~.
\end{eqnarray}  
Here we are writing the Wilson lines as shift vectors in the $\Gamma^{6,22}$
lattice. Thus, $U_1$ and $U_2$ are order two (${\bf Z}_2$) shifts. 
Here $e_1/2$ and $e_2/2$ are the right-moving shifts in $\Gamma^{2,2}$. The 
$\Gamma^{4,4}$ shifts are given by $P^{(1)}_R =P^{(1)}_L +E_1 +E_3$, 
$P^{(1)}_L =-h{\tilde E}^4$, and $P^{(2)}_R =P^{(2)}_L +E_2 +E_4$, 
$P^{(2)}_L =h{\tilde E}^3$. The $SO(32)$ shifts are given in the $SO(10)^3
\otimes SO(2)$ basis. In this basis, ${\bf 0}$ stands for the null vector, 
${\bf v}$($V$) is the vector weight, whereas
${\bf s}$($S$) and ${\overline {\bf s}}$(${\overline S}$) are the
spinor and anti-spinor weights of $SO(10)$($SO(2)$). (For $SO(2)$, $V=1$,
$S=1/2$ and ${\overline S}=-1/2$.) These Wilson lines 
break the gauge symmetry to $SU(3) \otimes R\otimes SO(10)^3 \otimes SO(2)$.
Note that $R$ is not affected. All the gauge bosons come from
the unshifted sector, whereas the shifted sectors give rise to massive states 
only. Note that, for each twist and/or shift in the model-building, 
we have implicitly chosen 
the spin structures of the right-moving world-sheet fermions to be 
compatible with the world-sheet supersymmetry.

{}Now we introduce the following ${\bf Z}_3$ twist on the $N1$ model:
\begin{equation}
 (\theta \vert\vert 0)(\Theta \vert\vert \Theta)({\cal P} \vert 2/3)~,
\end{equation}
where $\theta$ is a ${\bf Z}_3$ twist (that is, a $2\pi /3$ rotation) that acts
only on the right-moving part of the $\Gamma^{2,2}$ lattice (and the 
corresponding oscillator excitations), and the left-moving part is 
untouched. This is an asymmetric orbifold.
The $\Gamma^{4,4}$ lattice is twisted symmetrically by the 
${\bf Z}_3 \otimes {\bf Z}_3$ $\Theta$ twist.
The three $SO(10)$s are permuted by the action of the
${\bf Z}_3$ outer automorphism twist ${\cal P}$:
$\phi^I_1 \rightarrow
\phi^I_2 \rightarrow \phi^I_3 \rightarrow \phi^I_1$, where the real bosons
$\phi^I_p$, $I=1,...,5$, correspond to the $p^{\mbox{th}}$ $SO(10)$ 
subgroup, $p=1,2,3$. We can define new bosons $\varphi^I \equiv {1\over
\sqrt{3}}(\phi^I_1 +\phi^I_2 +\phi^I_3)$; the other ten real bosons are
complexified via linear combinations $\Phi^I \equiv {1\over
\sqrt{3}}(\phi^I_1 +\omega\phi^I_2 +\omega^2 \phi^I_3)$ and
$(\Phi^I)^\dagger \equiv {1\over \sqrt{3}}(\phi^I_1 +\omega^2\phi^I_2 
+\omega \phi^I_3)$, where $\omega =\exp(2\pi i /3)$.
Under ${\cal P}$, $\varphi^I$ is invariant, 
while $\Phi^I$ ($(\Phi^I)^\dagger$) are eigenstates with 
eigenvalue $\omega^2$ ($\omega$), {\em i.e.}, modded out. 
The ${\bf Z}_3$ invariant states form irreducible representations (irreps) 
of $SO(10)_3$.
Finally, string consistency requires the inclusion of the $2/3$ shift 
in the $SO(2)$ lattice.

{}The model (which we will refer to as $A1$) that results from twisting by 
the above ${\bf Z}_3$ twist has $N=1$ space-time supersymmetry. 
First, we discuss the untwisted sector of this model. All the gauge bosons  
come from the untwisted sector, and the gauge group is $SU(3)_1 \otimes R_1
\otimes SO(10)_3 \otimes U(1)$, where $R_1 \subset R$ depends on the value of
the modulus $h$. At the generic point $0<h<1$, $R_1 =U(1)^2$. At $h=0$,
$R_1$ is enhanced to $U(1)^4$, and for $h=1$, $R_1 =SU(3)_3$.
The latter case corresponds to a special
breaking $SO(8)_1 \supset SU(3)_3$ that results from the $\Theta$ twist.
This can be understood as a ${\bf Z}_3$ twist in the $SU(3)_1$
subgroup of $SO(8)_1\supset SU(3)_1 \otimes U(1)^2$ 
accompanied by a ${\bf Z}_3$ twist on the ${\bf Z}_3$ symmetry 
in the $U(1)^2$ subgroup, the latter simply 
being the triality symmetry of the $SO(8)_1$ Dynkin
diagram under which $8_v \rightarrow 8_s \rightarrow 8_c \rightarrow 8_v$.
Besides the gauge supermultiplets, other massless states appearing in 
the untwisted sector are $3$ copies of
adjoint Higgs fields of $SO(10)_3$. There are also 
$3$ copies of massless states in irreps of $R_1$. 
For example, at $h=1$, we have $3$ copies of chiral fermions in 
${\bf 10}_L$ of $SU(3)_3$ (here we define these states to be left-handed). 

{}The twisted sectors give rise to chiral matter fields of $SO(10)_3$. 
The asymmetric ${\bf Z}_3$ twist $(\theta \vert\vert 0)$
in $\Gamma^{2,2}$ contributes only a factor of $1$ to the number of 
fixed points as the factor $3$ contributed by the right movers is 
cancelled against the volume factor
of the invariant sublattice, which is $\Gamma^2$. Similarly, the outer
automorphism twist contributes only one fixed point. This follows from the 
form of the invariant sublattice, which is 
$\Gamma^6 =\{(\sqrt{3}{\bf q} \vert Q) \}$, where
$({\bf q} \vert Q)=({\bf 0} \vert 0), ({\bf v} \vert V), 
({\bf s} \vert S), ({\overline {\bf s}} \vert {\overline S})$. 
The momenta in the twisted and inverse twisted
sector belong to the shifted
dual lattices ${\tilde \Gamma}^6 +({\bf 0}\vert \pm 2/3)$, respectively, where
${\tilde \Gamma}^6 =\{({\bf q}/\sqrt{3} \vert Q) \}$. 
The only non-trivial contribution
to the number of fixed points in the twisted sectors comes from the symmetric
${\bf Z}_3$ twist $(\Theta \vert\vert \Theta)$ in $\Gamma^{4,4}$.
This twist contributes $9=3_R \times 3_L$ fixed points. 
So there are $9$ fixed points in the twisted sector. The left-moving fixed
points fall under irreps of the $R_1$ group. For $R_1=SU(3)_3$
we have three copies (due to the three right-moving fixed points) of 
massless states in the $SU(3)_1\otimes SU(3)_3 \otimes SO(10)_3$ irreps 
$({\bf 1}, {\overline {\bf 3}}, {\bf 16})(-1)_L$, 
$({\bf 1}, {\overline {\bf 3}},
{\bf 10})(+2)_L$, $({\bf 1}, {\overline {\bf 3}}, {\bf 1})(-4)_L$ 
(Here we give the $U(1)$ charge in parentheses, and its normalization 
is $1/6$). Note that the $SU(3)_3$ chiral anomaly in the twisted 
sectors is cancelled by that in the untwisted sector as a 
${\bf 10}_L$ of $SU(3)_3$ has $27$ times the anomaly contribution 
of a ${\bf 3}_L$. The model is also $U(1)$ anomaly-free
due to the underlying $E_6$ structure of the $SO(10)_3 \otimes U(1)$ 
matter fields as can
be seen from the branching ${\bf 27} ={\bf 16}(-1) +{\bf 10} (+2) +{\bf 1}
(-4)$ under the breaking $E_6 \supset SO(10) \otimes U(1)$. 

{}To obtain the final model, 
let us orbifold the $A1$ model by the following symmetric 
${\bf Z}_2$ twist:
\begin{equation}
 (0\vert\vert e_1/2)(-{\bf 1}\vert\vert -{\bf 1})(0^{15} \vert 0)~.
\end{equation}
Here the left-moving momenta of $\Gamma^{2,2}$
are shifted by $ e_1/2$, while
$\Gamma^{16}$ is untouched. $\Gamma^{4,4}$ is
symmetrically twisted by a diagonal ${\bf Z}_2$ twist (${\bf 1}$ 
is a $4\times 4$ identity matrix). This ${\bf Z}_2$ orbifold preserves the
$N=1$ supersymmetry. 

{}First, we discuss the untwisted sector $U$.
All the gauge bosons still come from the 
untwisted sector, and the gauge group is now $SU(2)_1 \otimes U(1) \otimes M
\otimes SO(10)_3 \otimes U(1)$. The $SU(2)_1 \otimes U(1)$ factor 
emerges from the regular breaking $SU(3)_1 \supset SU(2)_1 \otimes 
U(1)$, due to the $ e_1/2$ shift. Since this shift is required by 
string consistency, we see that $SU(2)_1$ is the biggest possible
hidden sector in our construction. 
$M$ is a subgroup of $R_1$, depending on the value of $h$.
For $0<h<1$, $M=U(1)$. For $h=0$, $M$ is
enhanced to $U(1)\otimes U(1)$. For $h=1$, $M$ is enhanced to $SU(2)_3 \otimes 
U(1)$, which is a result of the regular breaking $SU(3)_3 \supset SU(2)_3
\otimes U(1)$. Let us focus on the $h=1$ case, since the other cases 
are simpler. Note that there are no massless states in the
non-trivial irreps of $SU(2)_1 \otimes U(1) \otimes U(1)$. 
Two out of the three copies of the
massless states in the irreps of $R_1 \otimes SO(10)_3$ in the
$A1$ model have ${\bf Z}_2$ phase $-1$, whereas the third copy has the 
phase $1$. Since the adjoint irreps of $SO(10)_3$ are singlets under $R_1$,
only one copy of the massless 
$SO(10)_3$ adjoint Higgs fields remains in the final model. 
We also have one copy of ${\bf 1}(-6)$
and ${\bf 3}(0)$ each, and two copies of ${\bf 2}(-3)$ and ${\bf 4}(+3)$ each
(The $U(1)$ charge is normalized to $1/3\sqrt{2}$).
These states arise as a result of the branching (under $SU(3)_3 \supset
SU(2)_3 \otimes U(1)$) ${\bf 10}={\bf 1}(-6) + {\bf 2}(-3) +{\bf 3}(0) +
{\bf 4}(+3)$, where the singlet and the triplet have the ${\bf Z}_2$ phase $1$,
while the doublet and the quartet have the phase $-1$.

{}Next, let us consider the ${\bf Z}_3$ twisted (plus its inverse) 
sector $T3$. We start with the $9$ fixed points in this sector. 
Of these $9$ fixed points, the one at the origin is invariant under the
${\bf Z}_2$ twist. The remaining $8$ fixed points form $4$ pairs, and
the ${\bf Z}_2$ twist permutes the $2$ fixed points in each pair.
Forming $4$ symmetric and $4$ antisymmetric 
combinations, we have $9=5(1)+4(-1)$ (where the 
${\bf Z}_2$ phases are given in parentheses); that is, $5$ of the original
$9$ are invariant under the  ${\bf Z}_2$ twist.
Since there is no relative 
phase between the $T2$ and $T3$ sectors, these $5$ 
copies of the $SO(10)_3$ chiral matter fields survive, while the other 
$4$ are projected out.
These $5$ copies transform in the irreps of $M$. 
We have $2$ copies of $({\bf 1}, {\bf 2}, {\bf 16})
(0,-1,-1)_L$ and one copy of $({\bf 1}, {\bf 1}, {\bf 16})(0,+2,-1)_L$, plus
the corresponding  vector and singlet irreps of
$SO(10)_3$ (the $U(1)$ charges are normalized to 
$(1/\sqrt{6} ,1/3\sqrt{2}, 1/6)$).

{}Next, consider the ${\bf Z}_6$ twisted (plus its inverse) sector $T6$. 
The sublattice invariant
under the ${\bf Z}_6$ twist is the same as that for the ${\bf Z}_3$ twist.
The number of fixed points in the $T6$ sector is one. The massless chiral
fields are singlets under $SU(2)_1 \otimes SU(2)_3$:
$({\bf 1}, {\bf 1}, {\bf 16})(\pm 1, -1, -1)_R$, plus  the 
corresponding vector and singlet irreps of $SO(10)_3$. 
Note that these states are right-handed, and come in pairs ($\pm 1$ of the
first $U(1)$ charge). So, effectively, we have a total of $3=5-2$ 
left-handed chiral families of ${\bf 16}$'s.

{}Last, we consider the ${\bf Z}_2$ twisted sector $T2$. Let us 
consider first the twisted sector 
of the ${\bf Z}_2$ orbifold of the $N1$ model, and then 
its ${\bf Z}_3$-invariant states that are present in the final model. 
The sublattice 
invariant under the ${\bf Z}_2$ twist is given by the sublattice of
$\Gamma^{2,2}\otimes \Gamma^{16}$ invariant under the Wilson lines $U_1$ and 
$U_2$. The metric of this sublattice has determinant $16$. Therefore, the 
number of fixed points is $4_R \times 4_L /\sqrt{16} =2_R \times 2_L$. 
The ${\bf Z}_2$ orbifold breaks $SU(3) \otimes SO(8)$ to 
$SU(2) \otimes U(1) \otimes SU(2)^4$, with two massless sets of 
$({\bf 2}, {\bf 1}, {\bf 2}, {\bf 2},{\bf 2})(0)$ 
and $({\bf 1}, {\bf 2}, {\bf 1}, {\bf 1},{\bf 1})(\pm 3)$. 
Now consider the action of the ${\bf Z}_3$ twist.
It converts the last three $SU(2)$ to $SU(2)_3$, while 
breaking the second $SU(2)$ to $U(1)$. The resulting ${\bf Z}_3$-invariant 
massless states are $({\bf 2}, {\bf 2})$ and $({\bf 2}, {\bf 4})$
(in $SU(2)\otimes SU(2)_3$) plus a pair of singlets. 
All the states in non-trivial irreps of 
$SO(10)_3 \otimes U(1)$ are massive.
This concludes our construction. 

{}We have explored various combinations of ${\bf Z}_3$ 
twists and shifts, but failed to obtain the $3$-family feature. 
This leads us to the additional ${\bf Z}_2$ twist used above. 
Within this framework,
we have also obtained a variation of the 
above $M=SU(2)_3 \otimes U(1)$ model; the only difference is the 
assignment of the $M$ quantum numbers and $U(1)$ charges. 
These points and other GUSTs will be discussed in Ref.~\cite{zurab}.
In conclusion, we see that the realization of the $3$-family grand 
unification in string theory imposes very powerful constraints in the
moduli space. 

\acknowledgements
We thank Paul Aspinwall and Gary Shiu for discussions.
This work is supported in part by the National Science Foundation.

\begin{table}[t]
\begin{tabular}{|c|l|l|} \hline
 M & $SU(2)_3 \otimes U(1)$ & $U(1)\otimes U(1)$ \\ \hline
   & $ ({\bf 1},{\bf 1},{\bf 45})(0,0,0)$ &
 $ ({\bf 1},{\bf 45})(0,0,0,0)$ \\
   & $ ({\bf 1},{\bf 3},{\bf 1})(0,0,0)$ &
 $2 ({\bf 1},{\bf 1})(0,+12,0,0)_L$ \\
 U & $ ({\bf 1},{\bf 1},{\bf 1})(0,-6,0)_L$ &
 $2 ({\bf 1},{\bf 1})(0,0,+12,0)_L$ \\
   & $2 ({\bf 1},{\bf 4},{\bf 1})(0,+3,0)_L$ &
 $3 ({\bf 1},{\bf 1})(0,-6,0,0)_L$ \\
   & $2 ({\bf 1},{\bf 2},{\bf 1})(0,-3,0)_L$ &
 $3 ({\bf 1},{\bf 1})(0,0,-6,0)_L$ \\ \hline
   & $2 ({\bf 1},{\bf 2},{\bf 16})(0,-{1},-{1})_L$ &
 $2 ({\bf 1},{\bf 16})(0,+{2},+{2},-1)_L$ \\
   & $2 ({\bf 1},{\bf 2},{\bf 10})(0,-{1},+2)_L$ &
 $2 ({\bf 1},{\bf 10})(0,+2,+2,+{2})_L$ \\
   & $2 ({\bf 1},{\bf 2},{\bf 1})(0,-{1},-{4})_L$ &
 $2 ({\bf 1},{\bf 1})(0,+{2},+{2},-4)_L$ \\
   & $ ({\bf 1},{\bf 1},{\bf 16})(0,+{2},-{1})_L$ &
 $ ({\bf 1},{\bf 16})(0,-{4},-{4},-{1})_L$ \\
   & $ ({\bf 1},{\bf 1},{\bf 10})(0,+{2},+{2})_L$ &
 $ ({\bf 1},{\bf 10})(0,-{4},-{4},+{2})_L$ \\
 T & $({\bf 1},{\bf 1},{\bf 1})(0,+{2},-{4})_L$ &
 $ ({\bf 1},{\bf 1})(0,-{4},-{4},-4)_L$ \\
 3 & &
 $ ({\bf 1},{\bf 16})(0,-{4},+{2},-{1})_L$ \\
   & &
 $ ({\bf 1},{\bf 10})(0,-{4},+{2},+{2})_L$ \\
   & &
 $ ({\bf 1},{\bf 1})(0,-{4},+2,-{4})_L$ \\
   & &
 $ ({\bf 1},{\bf 16})(0,+{2},-{4},-{1})_L$ \\
   & &
 $ ({\bf 1},{\bf 10})(0,+{2},-{4},+{2})_L$ \\
   & &
 $ ({\bf 1},{\bf 1})(0,+{2},-{4},-{4})_L$ \\
 \hline
  & & \\
 T & $ ({\bf 1},{\bf 1},{\overline {\bf 16}})
                                (\pm 1,+{1},+{1})_L$ &
 $ ({\bf 1},{\overline {\bf 16}})
                    (\pm 1,+{1},+{1},+{1})_L$ \\
 6 & $ ({\bf 1},{\bf 1},{\bf 10})(\pm 1,+{1},-{2})_L$ &
 $ ({\bf 1},{\bf 10})(\pm 1,+{1},+{1},-{2})_L$ \\
   & $({\bf 1},{\bf 1},{\bf 1})(\pm 1,+{1},+{4})_L$ &
 $ ({\bf 1},{\bf 1})(\pm 1,+{1},+{1},+{4})_L$ \\
 \hline
   &  $ ({\bf 2},{\bf 2},{\bf 1})(0,0,0)$ &
 $2 ({\bf 2},{\bf 1})(0,-{3},-{3},0)_L$ \\
 T &  $  ({\bf 2},{\bf 4},{\bf 1})(0,0,0)$ &
 $  ({\bf 2},{\bf 1})(0,\pm {9},+{3},0)_L$ \\
 2 & &
 $  ({\bf 2},{\bf 1})(0,+{3},\pm {9},0)_L$ \\
   &  $  ({\bf 1},{\bf 1},{\bf 1})(\pm 3,-3,0)_L$ &
 $ ({\bf 1},{\bf 1})(\pm 3,-{3},-{3},0)_L$ \\ \hline
\end{tabular}
\caption{The massless spectra of the two models
$SU(2)^2 \otimes SO(10)\otimes U(1)^3$
and $SU(2) \otimes SO(10) \otimes U(1)^4$.
The gravity, dilaton and gauge supermultiplets are not shown.}
\label{table}
\end{table}


\begin{references}

\bibitem[*]{foot1}E-mail address: zurab@hepth.cornell.edu. Address after
September 1, 1996: Lyman Laboratory, Harvard University, Cambridge, MA 02138.

\bibitem{lew}
D.C. Lewellen, Nucl. Phys. {\bf B337} (1990) 61.

\bibitem{schw}
J.A. Schwartz, Phys. Rev. {\bf D42} (1990) 1777.

\bibitem{orb}L. Dixon, J. Harvey, C. Vafa and E. Witten, Nucl. Phys.
{\bf B261} (1985) 620; {\bf B274} (1986) 285;
K.S. Narain, M.H. Sarmadi and C. Vafa, Nucl. Phys.
{\bf B288} (1987) 551.

\bibitem{erler} J. Erler, "Asymmetric Orbifolds and Higher Level Models",
SCIPP 96/10, hep-th/9602032.

\bibitem{try} A. Font, L.E. Ib\'a\~nez, and F. Quevedo,
Nucl. Phys. {\bf B345} (1990) 389;
S. Chaudhuri, S.-W. Chung, G. Hockney and J.D. Lykken,
Nucl. Phys. {\bf B456} (1995) 89;
G.B. Cleaver, Nucl. Phys. {\bf B456} (1995) 219;
A. Aldazabal, A. Font, L.E. Ib\'a\~nez and A.M. Uranga,
Nucl. Phys. {\bf B452} (1995) 3; hep-th/9508033.
See also L.E. Ib\'a\~nez, H.P. Nilles and F. Quevedo, Nucl. Phys.
{\bf B307} (1988) 109.

\bibitem{zurab}
Z. Kakushadze and S.-H.H. Tye, Cornell preprint CLNS-96/1413, hep-th/9607138.

\bibitem{narain} K.S. Narain, Phys. Lett. {\bf B169} (1986) 41;
K.S. Narain, M.H. Sarmadi and E. Witten, Nucl. Phys. {\bf B279} (1987) 369.

\end{references}
\end{document}